\documentclass[a4paper,11pt]{article}
\pdfoutput=1 
\usepackage{jheppub} 
\usepackage{etoolbox}
\patchcmd{\maketitle}{\@fpheader}{}{}{}

\usepackage[T1]{fontenc} 
\usepackage[utf8]{inputenc} 
\usepackage{microtype} 
\usepackage{lmodern} 
\usepackage{mdframed} 
\usepackage{graphicx}
\usepackage{todonotes} 

\usepackage{amsmath}
\usepackage{amsfonts}
\usepackage{amssymb}
\usepackage{stmaryrd}
\usepackage{mathtools}
\usepackage{mathrsfs}

\usepackage{mleftright} \mleftright
\usepackage{tensor} 
\usepackage{braket} 
\usepackage{mathrsfs} 

\usepackage{bbold}
\usepackage{color}
\usepackage{braket}
\usepackage{slashed}
\usepackage{hyperref}

\newcommand{\bc}{\textcolor{blue}}

\newcommand{\comment}[1]{}

\DeclareMathAlphabet{\mathfs}{U}{rsfs}{m}{n}                     %

\newcommand{\n}{\nonumber}
\newcommand{\be}{\nopagebreak[3]\begin{equation}}
\newcommand{\ee}{\end{equation}}
\newcommand{\bee}{\nopagebreak[3]\begin{equation*}}
\newcommand{\eee}{\end{equation*}}
\newcommand{\ba}{\nopagebreak[3]\begin{eqnarray}}
\newcommand{\ea}{\end{eqnarray}}
\newcommand{\baa}{\nopagebreak[3]\begin{eqnarray*}}
\newcommand{\eaa}{\end{eqnarray*}}

\title{Non-Relativistic Maxwell Chern-Simons Gravity}
\author[1,2]{Luis Avil\'es,}
\author[1]{Ernesto Frodden,}
\author[1,3,4]{Joaquim Gomis,}
\author[1,2]{Diego Hidalgo,}
\author[1]{Jorge Zanelli}

\affiliation[1]{Centro de Estudios Cient\'ificos (CECs), Av. Arturo Prat 514, Valdivia, Chile,}
\affiliation[2]{Departamento de F\'isica, Universidad de Concepci\'on, Casilla 160-C, Concepci\'on, Chile,}
\affiliation[3]{Instituto de Ciencias F\'isicas y Matem\'aticas, Universidad Austral de Chile, Valdivia, Chile,}
\affiliation[4]{Departament de F\'isica Qu\`{a}ntica i Astrof\'isica and Institut de Ci\`{e}ncies del Cosmos (ICCUB), Universitat de Barcelona, Mart\'i i Franqu\`{e}s 1, E-08028 Barcelona, Spain}
\emailAdd{aviles@cecs.cl}
\emailAdd{dhidalgo@cecs.cl}
\emailAdd{z@cecs.cl}
\preprint{CECS-PHY-17/05, ICCUB-18-001}

\abstract{We consider a non-relativistic (\textbf{NR}) limit of $(2+1)$-dimensional Maxwell Chern-Simons (\textbf{CS}) gravity with gauge algebra [Maxwell] $\oplus \ u(1)\oplus u(1)$. We obtain a finite NR  CS gravity with a degenerate invariant bilinear form. We find two ways out of this difficulty: To consider  i) [Maxwell] $\oplus\ u(1)$, which does not contain Extended Bargmann gravity (\textbf{EBG}); or, ii) the NR limit of  [Maxwell] $\oplus\ u(1)\oplus u(1)\oplus u(1)$, which is a Maxwellian generalization of  the EBG.}

\begin{document}

\maketitle
\flushbottom

\section{Introduction}
\label{sec:introduction}

Non-relativistic (\textbf{NR}) holography has become an interesting tool to deal with strongly coupled condensed matter systems \cite{Sachdev,Liu}. In a situation in which the spacetime curvature is small, classical gravity could be a good approximation in the bulk. Therefore, the study of nonrelativistic gravities \cite{Cartan:1923zea,Havas:1964zza,Trautman,Datcour,Kuchar:1980tw,Duval:1984cj,Julia:1994bs,DePietri:1994je,DePietri:1994rd,Horava:2009uw, Andringa:2010it,Andringa:2012uz,Shroers,Christensen:2013lma,Bergshoeff:2014uea, BRZ, Bergshoeff-Rosseel,Obers-Hartong,Bergshoeff:2017dqq} is a subject that could be useful to understand non-relativistic coupled systems in the boundary.

NR theories is a vast world characterized by giving up Lorentz symmetry as a fundamental ingredient. This symmetry has prevailed as a key ingredient for fundamental theories and therefore an elegant way to formulate a NR system is by considering a limiting process from a relativistic one. In this sense a non-relativistic system is a \textit{sector} of a more fundamental theory, but there are several different prescriptions for taking this limit. One prescription could priviledge time together with one (or several) spatial directions, as in theories for extended objects living in many dimensions such as string theories, where these are natural options \cite{GO,Gomis:2004pw,BGN}. Each of those NR limits may have a physical interest of its own. For each one the structure of the theory, and therefore its physical consequences, change in different ways. 

In this work, the NR limit is one in which the speed of light is taken to infinity. In this process only time is a special direction and it is called the {\it NR particle limit} because it preserves the rotation group of point-like objects.

In flat spacetime the standard NR limit corresponds to the contraction of the Poincar\'e group into the Galilei group. Group deformations of this type can be systematized in general with the In\"on\"u-Wigner contractions of the group algebra.

A well-defined NR limit of a Lagrangian system can be framed as a regular contraction of the relativistic symmetry algebra preserving the number of generators while keeping the fields and the action finite. In the limit $c\to \infty$ there might appear infinities in the expansion of the original Lagrangian. An interesting aspect, recently found in the literature, is that a Lagrangian system with a finite NR limit may require an enlargement of the field content of the relativistic theory \cite{GO,BRZ,Bergshoeff-Rosseel,BCG}. Up to now a general method for the inclusion of extra fields, algebra generators and the new pieces of the Lagrangian for the starting theory, is not known. 

An additional feature of the NR limit is that the symplectic form of the NR theory might become degenerate, making some fields not determined by the field equations, thus reducing the number of dynamical fields. In the case of a Chern-Simons (\textbf{CS}) formulation in three dimensions, the non-degeneracy of the bilinear invariant trace of gauge generators, $\langle G_A, G_B \rangle$, implies the non-degeneracy of the symplectic form, which would ensure dynamically indeterminate fields in the NR theory.

A $(2+1)-$dimensional example of the potential degeneracy occurs in the contraction to obtain the Bargmann algebra from the Poincar\'e algebra with an extra Abelian generator \cite{Shroers}. Neither the Galilei nor the Bargmann algebra admit a non-degenerate bilinear form. However, in $2+1$ dimensions with the use of a second central extension \cite{Levy-Leblond}, in what is called extended or exotic Bargmann algebra, a non-degenerate bilinear form can be obtained and a NR CS formulation is found \cite{Bergshoeff-Rosseel,Obers-Hartong}. Therefore, two Abelian generators with their corresponding fields are required at the relativistic level\footnote{The use of two $U(1)$ factors in the symmetry group was also considered in \cite{Hartong:2017bwq} in relation to $AdS_3/CFT_2$.} and those fields are crucial in order to obtain a finite Lagrangian in the NR limit.

Here we study the NR limit of a three-dimensional gravity theory coupled to electromagnetism using an extension of the Poincar\'e algebra commonly known as the Maxwell algebra \cite{Schrader,Bacry-1970}. In this system, the generators of translations obey $[P_A, P_B]=Z_{AB}$, where the new generator $Z_{AB}$ transforms as an antisymmetric second rank Lorentz tensor and commutes with spacetime translations. A particle realization of this algebra describes the motion of a charge interacting with an electromagnetic field with group manifold coordinates $x^A$ and $\theta^{AB}$, conjugate to $P_A$ and $Z_{AB}$, respectively.  In the Maxwell particle Lagrangian,  the components of the constant electromagnetic field $f_{AB}$ are the canonical momenta conjugate to $\theta^{AB}$ \cite{BG,Gomis:2017cmt}. This is the starting point of a rich family of extensions allowed by the Poincar\'e algebra, including higher rank tensors (for a recent classification see \cite{Gomis:2017cmt}). The physical relevance of these algebras is related to the motion of a charge distribution described by the coordinates of the center of mass and higher multipolar moments. These moments can be identified with the duals to the generators of the extended Poincar\'e algebras \cite{Gomis:2017cmt}.

A realization of the Maxwell algebra in gravity theories has been studied in \cite{dA-K-L,S-S,Hoseinzadeh:2014bla,Durka:2011nf}. There, the $Z_{AB}$ extension was used also in an attempt to include the cosmological constant, something that we do not do in this work. Instead, we would like to see the effect of including a covariantly constant electromagnetic field in the three-dimensional CS gravity, without introducing a cosmological constant, both in the relativistic and non-relativistic regimes. 

Note that as an extension of Poincar\'e symmetry, the Maxwell algebra is relativistic in the sense that temporal and spatial directions are on equal footing. How are these algebras modified in the limit $c\to \infty$ was discussed in \cite{Beckers:1983gp,Bonanos:2008kr}. Here we explore the NR limit of gravitation theories that admit a CS formulation using the Maxwell algebra in $2+1$ dimensions and we find several alternative CS theories for the NR Maxwell algebras. 

A NR theory with a finite Lagrangian is found by following the approach that leads to the exotic Bargmann algebra from the [Poincar\'e]$\ \oplus\ u(1)\oplus u(1)$ algebra. The addition of the two extra Abelian generators in the relativistic algebra is motivated by the existence of two central extensions of the Galilei algebra in 2+1 dimensions \cite{Levy-Leblond}. In the contraction of the Lagrangian this procedure yields two central extensions, one corresponds to the mass and the other to the non-commutativity of the boost generators \cite{Bergshoeff-Rosseel}. 

In our case, we consider the NR contraction of the [Maxwell]$\ \oplus\  u(1)\oplus u(1)$ algebra. The presence of the two Abelian generators is enough to guarantee a finite Lagrangian in the NR limit. In this case, however, the NR algebra has a degenerate bilinear form, which means that at least one of the NR fields is dynamically indeterminate. One way to circunvent this difficuty is to eliminate the Abelian generator associated to the non-commutativity of the NR boosts. This gives rise to a four-parameter family of NR Lagrangians. There is a particular choice of the free parameters in the NR limit of the theory that corresponds to the contraction of the [Maxwell]$\ \oplus\ u(1)$ algebra, where the Abelian gauge field is related to the mass central extension of the Galilei group. Both versions of NR Maxwell algebras can be seen to fit in the family of Galilean extended algebras constructed in \cite{Beckers:1983gp,Bonanos:2008kr}.

 An alternative way to circunvent the degeneracy difficulty found in the contraction of [Maxwell] $\ \oplus\ u(1)\oplus u(1)$ is to consider an extra $U(1)$ field at the relativistic level. Starting from the relativistic algebra [Maxwell]$\ \oplus\ u(1)\oplus u(1)\oplus u(1)$, we show that there is a generalization of the trasformation used in \cite{Bergshoeff-Rosseel} that includes the Maxwellian generators such that the resulting bilinear form is non degenerate. The NR CS action obtained through the contraction is what we call the Maxwellian Extended Bargmann Gravity, but it has also the Extended Bargmann Gravity and the Exotic Gravity as subcases.

\section{2+1 Relativistic Gravity and Maxwell algebra}
\label{relativistic-preliminares} 
\subsection{Chern-Simons action}\label{rela-preliminares} 

In this section, we construct a gauge quasi-invariant gravity action under the Maxwell algebra in $2+1$ spacetime dimensions. The Maxwell algebra consists of the nine generators: Spacetime rotations $J_A$,  spacetime translations $P_A$, and a new type of generators $Z_A$ characterized and introduced in \cite{Bacry-1970,Schrader}.  The non vanishing commutators among these generators are
\begin{eqnarray}\label{algebrarelativista}
\nonumber && [J_A,J_B]=\epsilon_{ABC} J^C, \quad \quad  [J_A,P_B]= \epsilon_{ABC}P^C, \\
&& [J_A,Z_B] = \epsilon_{ABC} Z^C, \quad \quad  [P_A, P_B] =  \epsilon_{ABC}Z^C ,
\end{eqnarray}
where the latin indeces are raised and lowered with the Minkowski metric, $\eta^{AB}= \text{diag}(-1,1,1)=\eta_{AB}$, and they are split as $A=(0,a)$, with $a=1,2$. The conventions for the Levi-Civita symbol are $\epsilon_{012}=1, \epsilon^{012}=-1$. The generators are in their dualized form $Z^A = \frac{1}{2} \epsilon^{ABC}Z_{BC}$, and $J^{A}= \frac{1}{2} \epsilon^{ABC}J_{BC}$, and their inverse forms are $J_{AB}=\epsilon_{ABC}J^C$. In \cite{dA-K-L,S-S,Hoseinzadeh:2014bla,Durka:2011nf} the $Z_{AB}$ was defined as $\Lambda \tilde{Z}_{AB}$ where $\Lambda$ is the cosmological constant. Here, instead, we would like to see the effect of including a covariant constant electromagnetic field in the three-dimensional CS gravitational system without trying to introduce a cosmological constant.

In order to construct the relativistic action we will consider the most general bilinear form
\begin{eqnarray}\label{relativistic-pairing}
&&\langle J_A, Z_B \rangle = \langle P_A, P_B \rangle =  \alpha_1 \eta_{AB}, \quad \langle J_A, J_B \rangle =  \alpha_2 \eta_{AB}, \quad \langle P_A, J_B \rangle =  \alpha_3 \eta_{AB},
\end{eqnarray}
where $\alpha_i$ are real arbitrary constants. The invariance of this bilinear form under the action of the Maxwell algebra requires that $\langle J_A, Z_B \rangle$ and $\langle P_A, P_B \rangle$ have the same global coefficient. Hence, the most general quadratic Casimir invariant is $C=\alpha_1( P^AP_A+J^A Z_A)+\alpha_2J^AJ_A+\alpha_3 P^AJ_A$.\\

A gauge-invariant gravity action with the $2+1$ Maxwell algebra can be constructed using the connection one-form $A=A_\mu dx^\mu$ taking values in the Maxwell algebra generated by $\{P_A, J_A, Z_A\}$, 
\begin{equation}\label{connection-1}
A = E^{B} P_B + W^{B} J_B + K^{B} Z_B,
\end{equation} 
where $E^B$, $W^B$, and $K^B$ are one-form fields. The CS form for this connection constructed with the invariant bilinear form (\ref{relativistic-pairing}) defines an action for the relativistic gauge theory for the Maxwell symmetry as
\begin{equation}\label{actionCS}
S_{R}=\int \langle A \wedge dA + \frac{1}{3} A\wedge[A,A] \rangle .
\end{equation}
The explicit form in terms of the one-form fields reads\footnote{Here, the wedge product $\wedge$ between differential forms is understood, i.e., $W^A E^B=W^{A}\wedge E^{B}$.}
\begin{multline}\label{LagR2}
 S_{R}  = \int   \left\{ \alpha_1\left(2K^A R_A(W) + E^AT_A\right)\right. +  \alpha_2\left(W^AdW_A\right. + \frac{1}{3}\left.\epsilon_{ABC}W^A W^B W^C\right)  \\
 \left. +\alpha_3E^A R_A(W) \right\},
\end{multline}
where each term accompanying $\alpha_i$ in the Lagrangian is quasi-invariant under the Maxwell symmetry. The Lorentz curvature, the torsion, and the $K^A$ curvature are given by
\begin{eqnarray}
R^A (W) & = & dW^A - \frac{1}{2}\epsilon^{ABC} W_B W_C, \\
R^A(E)  &= & D_WE^A=T^A,\\
R^A(K)&=&D_W K^A-\frac{1}{2}\epsilon^{ABC}E_BE_C,
\end{eqnarray}
where the covariant derivative is $D_W \Phi^A:= d\Phi^A-\epsilon^{ABC}W_B\Phi_C$.

The field equations derived from \eqref{LagR2} are
\begin{eqnarray} \label{eq-rel}
&&\delta E^A: \quad \quad 2\alpha_1 T_A +  \alpha_3 R_A(W)  =  0, \\
&&\delta W^A: \quad \quad 2\alpha_1 R_A(K) + 2\alpha_2 R_A(W) +\alpha_3 T_A  =  0 , \\
&&\delta K^A : \quad \quad 2 \alpha_1 R_A(W )  =  0.
\end{eqnarray}
{Clearly, these equations dynamically determine every field by the vanishing of every curvature $R_A(W)=0,$ $T_A =0$, and $R_A(K)=0$, provided $\alpha_1\neq 0$ (regardless of the choices for $\alpha_2$ and $\alpha_3$), otherwise $K^A$ would be a redundant field of the theory, i.e., and the bilinear form  (\ref{relativistic-pairing}) would be degenerate.

The vanishing of the $R_A(K)$ can be rephrased as the constancy of the covariant derivative of $K^A$: $2D_WK^A=\epsilon^{ABC}E_B E_C$. This is analogous to the constancy of the background electromagnetic field in flat space-time, considered to define a system invariant under the Maxwell symmetry. 

A second order formulation of this system could be considered by postulating $T^A=0$ and algebraicly solving for $W^A$ as a function of $E^A$ and $\partial E^A$, so that (\ref{eq-rel}) becomes a set of second order equations for the metric $g_{\mu \nu}=\eta_{AB}E^A_\mu E^B_\nu$ and $K^A$. This procedure would give rise to new degrees of freedom, analogous to the topologically massive gravity \cite{DJT,Wise}.

\subsection{Relativistic Symmetries and $U(1)$ Enlargements} 

An infinitesimal gauge transformation for the $\text{Maxwell}$ algebra-valued one-form connection $A$, is given by $\delta_\Lambda A =d \Lambda + [A, \Lambda]$, where $\Lambda= \rho^A P_A + \lambda^AJ_A + \Theta^A Z_A  $ is an algebra-valued gauge parameter. The infinitesimal transformation on the field components is
\begin{eqnarray}
 \delta_\Lambda E^A &=& D_W \rho^A - \epsilon^{ABC}E_B\lambda_C, \label{algebra-relativista1}\\
\delta_\Lambda W^A & = & D_W \lambda^A, \\
\delta_\Lambda K^A & = & D_W \Theta^A -\epsilon^{ABC} E_B \rho_C - \epsilon^{ABC}K_B \lambda_C. \label{algebra-relativista3}
\end{eqnarray} 

The relativistic CS action \eqref{LagR2} is invariant up to a boundary term under the infinitesimal transformations (\ref{algebra-relativista1})-(\ref{algebra-relativista3}).

As explained in the introduction, a straightforward limit of the relativistic action \eqref{LagR2} gives an infinite result. In order to cancel the divergence, one can include extra auxiliary Abelian fields. This choice is directly inspired by the [Poincar\'e]$\ \oplus\ u(1)\oplus u(1)$ algebra  that allows us to obtain the Extended Bargmann algebra and a CS theory for it \cite{Bergshoeff-Rosseel}. Following this pattern, we include two extra $U(1)$ one-form gauge fields, $M$ and $S$ in the connection \eqref{connection-1} as
\begin{equation}
A = E^{B}P_B + W^{B} J_B + K^{B} Z_B + M Y_1+SY_2.
\end{equation}

The invariant bilinear form for the new algebra, [Maxwell]$\oplus u(1)\oplus u(1)$, is a simple extension of the bilinear form of the original Maxwell algebra. The new elements can always be brought to satisfy $\langle Y_1, Y_1 \rangle = \alpha_4$, $\langle Y_1, Y_2 \rangle = \alpha_5$, and $\langle Y_2, Y_2 \rangle = 0$; with $\alpha_4$ and $\alpha_5$ arbitrary real constants. With this new connection and invariant bilinear form, the relativistic CS action \eqref{LagR2} becomes
\begin{multline}\label{LagR2U(1)}
S_{R}  = \int   \left\{ \alpha_1\left(2K^A R_A(W) + E^AT_A\right)\right. +  \alpha_2\left(W^AdW_A\right. + \frac{1}{3}\left.\epsilon_{ABC}W^A W^B W^C\right)  \\
\left. +\alpha_3E^A R_A(W) + \alpha_4 MdM + 2\alpha_5 MdS\right\}.
\end{multline}

As we show in the next section, the insertion of these gauge vector fields eliminates the divergences that arise in the contraction procedure.

\section{Non-Relativistic Maxwell Gravities}
\label{NR Maxwell Gravity} 
We now consider Inönü-Wigner contractions for the extended Maxwell algebra. In order to carry out the contractions we express the relativistic algebra generators with a linear combination of new generators that involves a dimensionless parameter $\xi$. By taking the limit $\xi \rightarrow \infty$ one obtains a NR version of the Maxwell algebra. As stated in the introduction, there are several inequivalent Inönü-Wigner contractions that may define different NR algebras. These new algebras can be used to construct new ($2+1$)-dimensional CS theories, the NR Maxwell gravities.

The inclusion of the extra $U(1)$ fields is essential for the limiting process, and it is through a suitable choice of the constants $\alpha_i$ that we obtain different meaningful NR versions of the Maxwell relativistic action.

In subsection \ref{section-NR1} we build a first non-relativistic Maxwell gravity as a CS theory for a connection valued on a NR version of the [Maxwell]$\ \oplus\  u(1)\oplus u(1)$ algebra. Because it is built in a similar way as the so-called Exotic Galilei algebra, we call it the Exotic Non-Relativistic Maxwell algebra (ENRM). It turns our that the dynamics of the theory is not fully determined because this algebra has a degenerate bilinear form. 

In subsection \ref{section-NR2}, we consider a contraction from the [Maxwell]$\ \oplus\ u(1)$ algebra. The resulting NR algebra admits a non-degenerate bilinear form, and consequently a dynamically well-defined CS theory. Because it requires the addition of a central extension, as in the Bargmann algebra with respect to the Galilei algebra, we call this second new algebra as the Bargmann Non-Relativistic Maxwell algebra (BNRM).

Finally, in subsection \ref{section-NR3}, we consider a contraction from the [Maxwell]$\ \oplus\ u(1)\oplus u(1) \oplus u(1)$ algebra. As the previous case, the resulting NR algebra admits a non degenerate bilinear form, and consequently a dynamically well-defined CS theory. The difference is that in this case the NR CS action has the Exotic Bargmann gravity as a subcase. We name the resulting NR theory as the Maxwellian Exotic Bargman gravity.
 
\subsection{Exotic Non-Relativistic Maxwell Algebra}\label{section-NR1} 

A NR version of the Maxwell algebra can be obtained from an Inönü-Wigner contraction of the relativistic[ [Maxwell]$\oplus u(1)\oplus u(1)$ algebra. The contraction can be motivated by considering the NR limit for the action of a massive Maxwell particle \cite{BG} and the transformations used  to obtain the extended Bargmann algebra in \cite{Bergshoeff-Rosseel}. The explicit relation between the relativistic generators with the new NR generators $\{\tilde{H}, \tilde{P}_a,\tilde{M},\tilde{J},\tilde{Z},\tilde{Z}_a,\tilde{G}_a, \tilde{S}\}$  is
\begin{eqnarray}\label{Full-contraction}
&& P_0= \frac{\tilde{H}}{2\xi}+ \xi \tilde{M},\quad J_0 = \frac{\tilde{J}}{2} + \xi^2 \tilde{S},\quad Z_a= \frac{1}{\xi} \tilde{Z_a},\quad Y_1=\frac{\tilde{H}}{2\xi}-\xi \tilde{M},\n \\
&& 
P_a = \tilde{P}_a,   \quad \hspace{1.3cm}J_a= \xi \tilde{G}_a, \quad \quad  \quad Z_0 = \tilde{Z}, \quad \quad   Y_2 =\frac{\tilde{J}}{2} - \xi^2 \tilde{S},
\end{eqnarray}
with $a=1,2$, $\epsilon_{ab}\equiv \epsilon_{0ab}$, $\epsilon^{ab}\equiv\epsilon^{0ab}$ such that $\epsilon_{ab}\epsilon^{ac}=-\delta^b_c$. The dimensionless parameter $\xi$ introduced to perform the contraction. We use a tilde for the non-relativistic generators. In the limit $\xi\rightarrow \infty$, the contracted algebra from \eqref{algebrarelativista} has the following non-zero commutators
\begin{eqnarray} \label{Full-NR-algebra}
\nonumber  [\tilde{G}_a,\tilde{P}_b] &=& - \epsilon_{ab}\tilde{M}, \quad \quad \quad  [\tilde{G}_a, \tilde{Z}_b] =- \epsilon_{ab}\tilde{Z}, \\ 
\nonumber  [\tilde{H},\tilde{G}_a]  &=&   \epsilon_{ab}\tilde{P}_b , \quad \quad \quad \quad [\tilde{J}, \tilde{Z}_a] =  \epsilon_{ab}\tilde{Z}_b, \\
\nonumber  [\tilde{J},\tilde{P}_a]  &=&   \epsilon_{ab}\tilde{P}_b , \quad \quad \quad \quad [\tilde{H},\tilde{P}_a] =  \epsilon_{ab}\tilde{Z}_b, \\
\nonumber  [\tilde{G}_a,\tilde{G}_b] &=& - \epsilon_{ab}\tilde{S}, \quad \quad \quad [\tilde{P}_a,\tilde{P}_b] =- \epsilon_{ab}\tilde{Z},
\\
\text{$\big[\tilde J,\tilde G_a\big]$}  &=&  \epsilon_{ab}\tilde G_b.
\end{eqnarray}

This algebra has three central extensions given by the generators $\tilde M$, $\tilde S$, and $\tilde Z$. Two of the central extensions are related to the two extra $U(1)$ generators, but the third comes directly from the Maxwell generator $Z_0$. Note that the Exotic Bargmann algebra in \cite{Bergshoeff-Rosseel} is the subalgebra obtained by suppressing the Maxwell generators $\tilde Z_a$ and $\tilde Z$ (they appear only on the right hand side).

The following non-relativistic (degenerate) invariant bilinear form is obtained directly from the contraction \eqref{Full-contraction} of the relativistic bilinear form \eqref{relativistic-pairing} 
\begin{eqnarray}\label{pairing-NR}
\nonumber	&& \langle \tilde{M}, \tilde{H} \rangle=\langle \tilde{J}, \tilde{Z} \rangle =-\tilde{\alpha}_1, \\
\nonumber	&& \langle \tilde{P}_a , \tilde{P}_b \rangle =\langle \tilde{G}_a, \tilde{Z}_b \rangle = \tilde{\alpha}_1 \delta_{ab}, \\
\nonumber && 	\langle \tilde{J}, \tilde{J} \rangle =-\tilde{\alpha}_2, \\
\nonumber	&& \langle \tilde{H}, \tilde{S} \rangle=\langle \tilde{M}, \tilde{J} \rangle =-\tilde\alpha_3,\\
	&&	 \langle \tilde{G}_a, \tilde{P}_b \rangle = \tilde\alpha_3\delta_{ab}, 
\end{eqnarray}
where $\tilde\alpha_1=\alpha_1$, $\tilde\alpha_2=\alpha_2$, and $\tilde\alpha_3=\alpha_3/\xi$ are taken as constants in the NR theory. The coefficients $\alpha_4$ and $\alpha_5$ are also present in the previous bilinear form, but we need to choose them as  $\alpha_4=\alpha_1$ $\alpha_5=\alpha_3$. As we will see in the following, these particular values guarantee finiteness of the NR Lagrangian.  While invariant under the NR algebra, the bilinear form (\ref{pairing-NR}) is degenerate. In fact, the ENRM algebra \eqref{Full-NR-algebra} can not be equipped with a non-degenerate invariant bilinear form. To prove this one may solve the most general bilinear form that is invariant under the non-relativistic algebra (\ref{Full-NR-algebra}) and check that its determinant vanishes. 

In order to build a CS action invariant up to a surface term under the algebra \eqref{Full-NR-algebra}, we consider the one-form connection
\begin{equation}\label{NR-connection}
\tilde{A} = \tau \tilde{H} + e^{a} \tilde{P}_a + \omega \tilde{J} + \omega^{a} \tilde{G}_a  + k \tilde{Z} + k^a \tilde{Z}_a+m \tilde{M} +s\tilde S.
\end{equation}

The NR fields are related to the relativistic ones by the inverse of the transformation \eqref{Full-contraction} in order to ensure that $A=\tilde A$ \cite{BRZ}. The curvature associated to this connection is 
\begin{equation}\label{NR-curvatures}
\tilde{F}(\tilde A) = R(\tau) \tilde{H} + R^a(e^{b}) \tilde{P}_a + R(\omega) \tilde{J} + R^a(\omega^{b}) \tilde{G}_a  + R(k) \tilde{Z} + R^a(k^b) \tilde{Z}_a+R(m) \tilde{M} +R(s)\tilde S,
\end{equation}
where we have written it in terms of the field curvatures
\begin{eqnarray}
\nonumber R(\tau ) & = & d \tau, \\
\nonumber R^a(e^b) & =  & de^a + \epsilon^{ac}\omega  e_c + \epsilon^{ac} \tau \omega_c  , \\
\nonumber R(\omega) & = & d \omega, \\
\nonumber R^a(\omega^b) & = & d\omega^a + \epsilon^{ac} \omega \omega_c,\\
\nonumber R(k) & = & dk + \epsilon^{ac} \omega_a k_c + \frac{1}{2} \epsilon^{ac} e_a e_c  ,\\
\nonumber R^a(k^b) & = & dk^a + \epsilon^{ac} \omega k_c  + \epsilon^{ac} \tau e_c  ,\\
\nonumber R(m) & = & dm + \epsilon^{ac} e_a \omega_c ,\\
R(s) & = & ds + \frac{1}{2}\epsilon^{ac}\omega_a \omega_c.
\end{eqnarray}

These curvatures are covariant under the ENRM algebra \eqref{Full-NR-algebra}. We can summarize the non-relativistic notations in the following table:

\begin{table}[h]
	\centering
	\begin{tabular}
		{ |p{3.8cm}||p{2cm}|p{2cm}|p{2cm}|}
		\hline
		\multicolumn{4}{|c|}{Non-relativistic elements} \\
		\hline
		Symmetry & Generators &Gauge fields & Curvature \\
		\hline
		\hline 
		Time translations   & $\tilde{H}$    &   $\tau$ & $R(\tau)$\\
		Space translations& $\tilde{P}^a$&  $e^a$   & $R_a(e^b)$\\
		Boosts &$\tilde{G}^a$ & $\omega^a$&  $R_a(\omega^b)$\\
		Spatial rotations    & $\tilde{J}$&$\omega$  &  $R(\omega)$\\
		Maxwell central charge &  $\tilde{Z}$   & $k$& $R(k)$\\
		Maxwell spatial field &  $\tilde{Z}^{a}$   & $k^a$&  $R_a(k^b)$\\
		First central charge & $\tilde{M}$  & $m$   & $R(m)$\\
		Second central charge & $\tilde{S}$  & $s$   & $R(s)$\\
		\hline
	\end{tabular}
\end{table}

The NR CS action \eqref{actionCS} built from the connection \eqref{NR-connection} and the bilinear form \eqref{pairing-NR} can be written as
\begin{eqnarray}\label{Full-NRaction}
\nonumber  S_{NR}& =& \int  \tilde{\alpha}_1 \left[-2k R(\omega)+2k^a R_a(\omega^b)+e^aR_a(e^b)-\tau R(m)-mR(\tau)  \right]
 -\tilde{\alpha}_2\omega R(\omega)  \\
&&+ \tilde{\alpha}_3\left[ e^aR_a(\omega^b) -\tau R(s)-m R(\omega) \right].
\end{eqnarray}

It is worth noting that, as expected, for the choices $\tilde{\alpha}_1=0=\tilde{\alpha}_2$ one obtains the same NR action derived in \cite{Bergshoeff-Rosseel}, which is a CS action for the Exotic Bargmann algebra.

The field equations are
\begin{eqnarray}\label{degEOM}
\delta \tau : && \quad  \tilde{\alpha}_1R(m)+\tilde{\alpha}_3 R(s)  =  0, \\
\nonumber \delta e^a : &&\quad  \tilde{\alpha}_1R_a(e^b)+\tilde{\alpha}_3 R_a(\omega^b)  =  0, \\
\nonumber \delta \omega : &&\quad  \tilde{\alpha}_1(2R(\omega)+R(k))+\tilde{\alpha}_3 R(m)+\tilde{\alpha}_2 R(\omega)  =  0, \\
\nonumber \delta \omega^a : && \quad  \tilde{\alpha}_1R_a(k^b)+\tilde{\alpha}_3 R_a(e^b)  =  0, \\
\nonumber \delta k :&& \quad \tilde{\alpha}_1 R(\omega)  =  0, \\
\nonumber \delta k^a : && \quad \tilde{\alpha}_1 R_a(\omega^b) =  0, \\
\nonumber \delta m : && \quad \tilde{\alpha}_1 R(\tau)+\tilde{\alpha}_3 R(\omega)  =  0, \\
\nonumber \delta s : && \quad \tilde{\alpha}_3 R(\tau	)  =  0,
\end{eqnarray}
which implies $R(\tau)=R(\omega)=0=R_a(\omega^b)=R_a(e^b)=R_a(k^b)$, and
\begin{equation}\label{R=R=R}
 R(k)= (\tilde{\alpha}_3 / \tilde{\alpha}_1)^2R(s) = -(\tilde{\alpha}_3 / \tilde{\alpha}_1)R(m). 
 \end{equation} 

From the last equations it is apparent that one of the three curvatures $R(m)$, $R(s)$, $R(k)$ is arbitrary, which is a consequence of the degenerate bilinear form (see Appendix A). 
 
On the other hand, since $A=\tilde A$, the components of the relativistic gauge fields in terms of the NR ones can be expressed as follows 
\begin{eqnarray}
\nonumber && E^{0} = \xi \tau + \frac{1}{2\xi} m, \hspace{1.0cm}  W^0 =  \omega+ \frac{1}{2 \xi^2}s, \hspace{1.0cm} W^a  =  \frac{1}{\xi}w^a, \hspace{1.0cm} E^a=e^a,\\
&& M = \xi \tau - \frac{1}{2\xi} m, \hspace{1.3cm} S=   \omega- \frac{1}{2 \xi^2}s, \hspace{1.1cm} K^a  =  \xi k^a, \hspace{1.2cm} K^0=k.
\label{transfields}
\end{eqnarray}

Using these last expressions on the action \eqref{LagR2} and then taking the limit $\xi\rightarrow \infty$, the action \eqref{Full-NRaction} is also obtained. This procedure gives in general a divergent piece for the NR action. Note however, that in our case there is a delicate balance between the extra $U(1)$ gauge fields and the coefficients $\alpha_i$ that exactly compensates the two divergences coming from the terms $-\xi^2 \tau d \tau$ and $-\xi\tau d\omega$, this is the main reason for using the extra fields. 

Now, in spite of the introduction of the two $U(1)$ fields to eliminate the divergences in the Lagrangian, the contraction process gave rise to an algebra with a degenerate bilinear form that does not produce a CS action with fully determinate dynamics. Then, a preliminary lesson is that guaranteeing finiteness of the action in the contraction process does not guarantee a non-degenerate bilinear form. Therefore a natural question emerges: Is it possible to have a different contraction yielding a NR algebra with a non-degenerate bilinear form?

From (\ref{R=R=R}), it is natural to expect that a solution of the degeneracy problem could come from eliminating one of the three one-form fields $m$, $k$ or $s$, corresponding to the three central extension of the algebra (\ref{Full-NR-algebra}). However, eliminating $m$ or $k$ by dropping from the beginning the associated central charges $\tilde M$ or $\tilde Z$ does not solves the problem because the resulting algebra does not admit a non-degenerate invariant bilinear form. On the other hand if one eliminates $\tilde S$ the resulting algebra admits a non-degenerate invariant bilinear form. This later option has a CS formulation and it is explored on the next subsection.

An alternative point of view would be to consider the NR Lagrangian (\ref{Full-NRaction}) as the starting point and  drop directly from it the fields $m$ or $k$. This would give rise to equations that determine all the remaining fields dynamically. However, there is not guarantee that the action would be gauge invariant under the resulting non-relativistic gauge symmetry because the Lagrangian would not necessarily be a CS form.

\subsection{Bargmann NR Maxwell Algebra}\label{section-NR2} 

At non-relativistic level, let us consider putting to zero the central generator $\tilde S$. Following the general steps shown in Appendix A we find the resulting algebra admits an invariant non-degenerate bilinear form
\begin{eqnarray}\label{pairing-NR'}
\nonumber	&& \langle \tilde{M}, \tilde{H} \rangle=\langle \tilde{J}, \tilde{Z} \rangle =-\tilde{\alpha}_1, \\
\nonumber	&& \langle \tilde{P}_a , \tilde{P}_b \rangle =\langle \tilde{G}_a, \tilde{Z}_b \rangle = \tilde{\alpha}_1 \delta_{ab}, \\
\nonumber && 	\langle \tilde{J}, \tilde{J} \rangle =-\tilde{\alpha}_2, \\
\nonumber	&& \langle \tilde{H}, \tilde{H} \rangle=\tilde\alpha_0,\\
	&&	 \langle \tilde{H}, \tilde{J} \rangle = \tilde\alpha_6, 
\end{eqnarray}
with $\tilde{\alpha}_1\neq 0$. The corresponding CS action is given by
\begin{eqnarray}\label{NewtonCartanMaxwell}
\nonumber  S_{NR}& =& \int  \tilde{\alpha}_1 \left[-2k R(\omega)+2k^a R_a(\omega^b)+e^aR_a(e^b)-\tau R(m)-mR(\tau)  \right]
 -\tilde{\alpha}_2\omega R(\omega)  \\
&&+ \tilde{\alpha}_0\tau R(\tau) + 2\tilde \alpha_6 \tau R(\omega).
\end{eqnarray}

The field equations are $R(\tau)=R(\omega)=R_a(\omega^b)=R_a(e^b)=R_a(k^b)=R(k)=R(m)=0$. In this way we have obtained an action invariant under the Bargmann non-relativistic Maxwell algebra that is also dynamically determinate. 

The previous theory is built out of the \bc{NR} algebra (\ref{Full-NR-algebra}) (with $\tilde S=0$). A remaining question is weather this theory can be obtained as a limiting process from a relativistic theory. We do not have a general answer. However, by considering $\alpha_2=0=\alpha_3$ from the beginning (thus $\tilde\alpha_2=0$ as well as $\tilde\alpha_0=\tilde \alpha_6$=0), this theory is obtained from a contraction of a [Maxwell]$\otimes U(1)$ relativistic theory. Because $\alpha_3=0$, there is not a divergence of the form $-\xi\tau d\omega$ in the limiting process, and therefore it is enough an unique $U(1)$ gauge field in order to eliminate the divergence coming from $\xi^2 \tau d\tau$. 

Then, we may take a minimalistic starting point at the very beginning by considering at the relativistic level only the term that guarantee a non-degenerated Maxwell CS gravity. That corresponds to use constants $\alpha_2=0=\alpha_3$ in the relativistic bilinear form \eqref{relativistic-pairing}. The Einstein-Hilbert and the Exotic gravity Lagrangian are absent from the beginning.

Then, we start with a relativistic algebra [Maxwell]$\otimes u(1)$ and consider the following transformation for the relativistic generators
\begin{eqnarray}\label{contraction}
&& P_0= \frac{\tilde{H}}{2\xi}+ \xi \tilde{M},\quad \quad \quad  J_0 = \tilde{J} ,\quad \quad \quad \quad  Z_a= \frac{1}{\xi} \tilde{Z_a},\quad \quad  Y_1= \frac{\tilde{H}}{2\xi}-\xi \tilde{M}, \\
&& 
P_a = \tilde{P}_a,   \quad \quad  \hspace{1.3cm}J_a= \xi \tilde{G}_a, \quad \quad  \quad \quad  Z_0 = \tilde{Z}.
\end{eqnarray}

Using these transformations we can obtain the Bargmann NR Maxwell algebra (BNRM) that has only two central extensions: $\tilde M$ and $\tilde Z$. The only change with respect to the algebra \eqref{Full-NR-algebra} is the vanishing of  $[\tilde{G}_a,\tilde{G}_b]=0$.

The BNRM algebra can be equipped with a non-degenerate invariant bilinear form
\begin{eqnarray}\label{pairing-NR2}
\nonumber	&& \langle \tilde{M}, \tilde{H} \rangle=\langle \tilde{J}, \tilde{Z} \rangle =-\tilde{\alpha}_1, \\
\nonumber	&& \langle \tilde{P}_a , \tilde{P}_b \rangle =\langle \tilde{G}_a, \tilde{Z}_b \rangle = \tilde{\alpha}_1 \delta_{ab}.
\end{eqnarray}

The NR connection is
\begin{equation}\label{NRconnection2}
\tilde{A} = \tau \tilde{H} + e^{a} \tilde{P}_a + \omega \tilde{J} + \omega^{a} \tilde{G}_a  + k \tilde{Z} + k^a \tilde{Z}_a+m \tilde{M}.
\end{equation}

Finally, the CS action invariant under the BNRM algebra is
\begin{eqnarray}\label{NR-action-2}
  S_{NR}  & =& \int \langle \tilde{A}d\tilde{A}   + \frac{1}{3}  [\tilde{A},\tilde{A}], \tilde{A} \rangle,  \\
& = &    \tilde{\alpha}_1\int  \left[-2k R(\omega)+2k^a R_a(\omega_b)+e^aR_a(e_b)-\tau R(m)-mR(\tau)  \right].
\end{eqnarray}

With this formulation the dynamics of the system is totally determined, that is $F(\tilde A)=0$, or equivalently, the field equations imply the vanishing of all the curvatures associated to the fields. That is, every curvature in \eqref{NR-curvatures} vanishes and $R(s)$ is absent in this set up.

The field equations for a CS theory imply that the connection is {\it locally} flat, $\tilde A=g^{-1}dg$, with $g$ any element of the gauge group (the connection is ``pure gauge''). Nevertheless, it is well-known that there are solutions which are not \textit{globally} flat due to non-trivial features of the space-time topology. Therefore, in looking for those solutions it is worth to analyze the equations of motion more carefully, component by component.

$\bullet$ Equation $R(\tau)=d\tau=0$ implies that the space-time can be foliated in an absolute time direction.

$\bullet$ At the relativistic level the vanishing of torsion, $T^A=0$, allows to solve algebraically for the connection in terms of the vielbein, its inverse and its first derivatives, $W^A=W^A(E^B, \partial E^B)$. This is what allows to pass from the first to the second order formulation of gravity. However, in the NR case the situation is more subtle. 

The relativistic vielbein $E^A$ is identified with the NR fields $\tau$ and $e^a$, while the relativistic connection $W^A$ is identified with the NR fields $\omega$ and $\omega^a$. However, by simply counting equations, the vanishing of $R_a(e^b)$ is not enough to solve for $\omega^a$ and $\omega$ in terms of the NR fields $(\tau,e^a)$. This is a known issue in the Newton-Cartan theory: the NR torsion-free equation only allows to solve the connection up to an arbitrary one-form. A natural way to fix the indeterminancy is to impose an extra equation for $\bar m$, a curvature equation $R(\bar m)$. In the present framework, the issue is automatically solved because the vielbein is actually identified with $\tau,$ $e^a$ {\it and} $m$, see (\ref{transfields}). And, in the NR theory $m$ satisfies the equation $R(m)=0$ \cite{Andringa:2010it}. Then, the system
\ba
R^a(e^b)&=&de^a + \epsilon^{ab} \tau \omega_b - \epsilon^{ab} e_b \omega =0,\\
R(m)&=&dm + \epsilon^{ab} e_a \omega_b=0,
\ea
is a linear system that can be algebraically solved for $(\omega^a,\omega)$ in terms of $(\tau,\ e^a, m)$. The explicit solutions are (see Appendix B)
\ba
\label{omega1}
\omega&=&\omega_\mu dx^{\mu}=\frac{1}{2}\left(
 \epsilon_{ab}\partial_{[\nu} e^{c}_{\rho]} e_{c\mu} e^{a\nu} e^{b\rho} 
-2\epsilon_{ab} \partial_{[\mu} e^{a}_{\nu]} e^{b\nu} - \tau_{\mu} \omega_{b\nu} e^{b\nu} \right)dx^\mu,\\
\label{omegaa}
\omega_a &=&\omega\indices{_a_\nu} dx^\nu=\left(\epsilon_{ab} \partial_{[\mu} m_{\nu]} e^{b \mu} + 2 \epsilon_{cd}e_{a\nu}\partial_{[\mu} e^{a}_{\xi]} e^{\xi b} \tau^\mu\right)dx^\nu.
\ea

Note that $\omega_{a\nu}$ should be replaced in the r.h.s. of (\ref{omega1}) to get the full expression for $\omega$. 

$\bullet$ Equation $R(\omega)=d\omega=0$ is trivially solved too. It means that the expression (\ref{omega1}) is locally a gradient of a scalar function. For example, if there is one missing point in the spatial section this allows for $\omega = d\theta$, which corresponds to the geometry of a cone in 2+1 dimensions. 

$\bullet$ Once the previous solutions are replaced, equation $R_a(\omega^b)=0$ is a second order differential equation for the fields $\tau,\ e^a$ and $m$. It corresponds to some of equations of the 2+1 Newton-Cartan equations of motion, i.e., a NR version of the Einstein field equations. 

$\bullet$ The remaining equations, $R_a(k^b)=0, R(k)=0$ determine dynamically the NR Maxwell fields.  

One interesting problem for future work would be to study the dynamical contents, and in particular the classical solutions of these theories both in the relativistic and NR regimes.

\subsection{Maxwellian Exotic Bargmann Gravity}\label{section-NR3} 

An alternative way to circunvent the degeneracy of the bilinear form in the [Maxwell]$\ \oplus\ u(1)\oplus u(1)$ system, is to add one more $u(1)$ gauge field. In this way one finds that there is a NR  contraction such that each term in the relativistic action has a NR counterpart. In other words, we do not need to assume the vanishing of $\alpha_2$ or $\alpha_3$ at the relativistic level. In particular the Einstein-Cartan term leads in the NR limit to a Newton-Cartan term. We observe that the final NR action contains three pieces: 1) the CS action for the exotic Bargmann algebra \cite{Bergshoeff-Rosseel,Obers-Hartong}; 2) the CS action for a new NR Maxwell algebra; and 3) a CS action for the NR exotic Gravity. The differences with respect to the previous NR Maxwell algebra presented in (\ref{Full-NR-algebra}), are the addition of the commutator  $[\tilde{Z}, \tilde{G}_a] =  \epsilon_{ab}\tilde{Z}_b$ plus a new central generator $\tilde T$ (in the r\^ole played previously by $\tilde Z$) in $[\tilde{G}_a, \tilde{Z}_b] =- \epsilon_{ab}\tilde{T}$  and $[\tilde{P}_a, \tilde{P}_b] =- \epsilon_{ab}\tilde{T}$.} Then, at the relativistic level we start with three $U(1)$ fields: $Y_1$, $Y_2$, and $Y_3$
\begin{equation}
A = E^{B}P_B + W^{B} J_B + K^{B} Z_B + M Y_1+SY_2+TY_3.
\end{equation}
The non zero entries for the bilinear form of the new Abelian generators are
\begin{eqnarray}\label{U3extra-pairing}
\nonumber	&& \langle Y_1, Y_1 \rangle=\langle Y_2, Y_3 \rangle =\alpha_1, \\
\nonumber && \langle Y_2, Y_2 \rangle ={\alpha}_2, \\
\nonumber	&& \langle Y_1, Y_2 \rangle=\alpha_3.
\end{eqnarray}

 The contraction is defined through the following identifications 
\begin{eqnarray}
&& P_0= \frac{\tilde{H}}{2\xi}+ \xi \tilde{M},\quad P_a = \tilde{P}_a,\quad Y_1=\frac{\tilde{H}}{2\xi}-\xi \tilde{M},\n \\
&& J_0 = \frac{\tilde{J}}{2} + \xi^2 \tilde{S},\quad J_a= \xi\tilde{G}_a, \quad Y_2 =\frac{\tilde{J}}{2} - \xi^2 \tilde{S}, \n \\
&& Z_0 = \frac{\tilde{Z}}{2\xi^2} + \tilde{T}, \quad Z_a= \frac{\tilde{Z}_a}{\xi},\quad Y_3=\frac{\tilde{Z}}{2\xi^2} - \tilde{T}.
\label{U3Full-contraction}
\end{eqnarray}

Note that the identifications for the Poincar\'e algebra generators (first two lines), the contracted algebra is the Exotic Bargmann algebra.

In terms of the NR generators and fields, the connection is 

\begin{equation}\label{U3NR-connection}
\tilde{A} = \tau \tilde{H} + e^{a} \tilde{P}_a + \omega \tilde{J} + \omega^{a} \tilde{G}_a  + k \tilde{Z} + k^a \tilde{Z}_a+m \tilde{M} +s\tilde S+t\tilde T.
\end{equation}
The contracted new NR Maxwell algebra, that we shall call {\it Maxwellian Exotic Bargmann} ({\bf MEB}) algebra , is
\begin{eqnarray} \label{U3Full-NR-algebra}
\nonumber  [\tilde{G}_a,\tilde{P}_b] &=& - \epsilon_{ab}\tilde{M}, \quad \quad \quad  [\tilde{G}_a, \tilde{Z}_b] =- \epsilon_{ab}\tilde{T}, \\ 
\nonumber  [\tilde{H},\tilde{G}_a]  &=&   \epsilon_{ab}\tilde{P}_b , \quad \quad \quad \quad [\tilde{J}, \tilde{Z}_a] =  \epsilon_{ab}\tilde{Z}_b, \\
\nonumber  [\tilde{J},\tilde{P}_a]  &=&   \epsilon_{ab}\tilde{P}_b , \quad \quad \quad \quad [\tilde{H},\tilde{P}_a] =  \epsilon_{ab}\tilde{Z}_b, \\
\nonumber  [\tilde{G}_a,\tilde{G}_b] &=& - \epsilon_{ab}\tilde{S}, \quad \quad \quad [\tilde{P}_a,\tilde{P}_b] =- \epsilon_{ab}\tilde{T},
\\
\text{$\big[\tilde J,\tilde G_a\big]$}  &=&  \epsilon_{ab}\tilde G_b, \quad \quad \quad \text{$\big[\tilde Z,\tilde G_a\big]$}  =  \epsilon_{ab}\tilde Z_b.
\end{eqnarray}
It is interesting to note that in contrast with the algebras found in \ref{section-NR1} and \ref{section-NR2},  this is not an extension of the Galilean algebra in the sense of \cite{Bonanos:2008kr}. The technical point is that the $\tilde Z$ generator does not appear as a central extension in any of the levels defined in \cite{Bonanos:2008kr}. Thus the expectation is that this algebra may be obtained by enlarging an algebra different from the Galilean one. 

The MEB algebra admits a non degenerate bilinear form obtained from the relativistic bilinear form (\ref{relativistic-pairing}) in the limit $\xi\to \infty$. To keep the action finite there is a rescaling of the relativistic action parameters that can be absorved on each non relativistic parameter as $\alpha_1=\tilde\alpha_1$, $\alpha_2={\xi^2}\tilde\alpha_2$, and $\alpha_3={\xi}\tilde\alpha_3$.  
\begin{eqnarray}\label{U3pairing-NR}
\nonumber	&& \langle \tilde{H}, \tilde{M} \rangle=\langle \tilde{J}, \tilde{T} \rangle =-\tilde{\alpha}_1, \\
\nonumber	&& \langle \tilde{P}_a , \tilde{P}_b \rangle =\langle \tilde{G}_a, \tilde{Z}_b \rangle = \tilde{\alpha}_1 \delta_{ab}, \\
\nonumber && \langle \tilde{J}, \tilde{S} \rangle =-\tilde{\alpha}_2, \\
\nonumber && \langle \tilde{G_a}, \tilde{G_b} \rangle =\tilde{\alpha}_2 \delta_{ab}, \\
\nonumber	&& \langle \tilde{H}, \tilde{S} \rangle=\langle \tilde{J}, \tilde{M} \rangle =-\tilde\alpha_3,\\
	&&	 \langle \tilde{G}_a, \tilde{P}_b \rangle = \tilde\alpha_3\delta_{ab}, 
\end{eqnarray}
which is non degenerate if $\tilde \alpha_1\neq 0$, in analogy with the relativistic case. 

The NR CS curvature along the generators of the MEB algebra is 
\begin{eqnarray}
\nonumber R(\tau ) & = & d \tau, \\
\nonumber R^a(e^b) & =  & de^a + \epsilon^{ac}\omega  e_c + \epsilon^{ac} \tau \omega_c  , \\
\nonumber R(\omega) & = & d \omega, \\
\nonumber R^a(\omega^b) & = & d\omega^a + \epsilon^{ac} \omega \omega_c,\\
\nonumber R(k) & = & dk,\\
\nonumber R^a(k^b) & = & dk^a + \epsilon^{ac} \omega k_c  + \epsilon^{ac} \tau e_c + \epsilon^{ac} k \omega_c ,\\
\nonumber R(m) & = & dm + \epsilon^{ac} e_a \omega_c ,\\
R(s) & = & ds + \frac{1}{2}\epsilon^{ac}\omega_a \omega_c\n\\
R(t) & = & dt + \epsilon^{ac} \omega_a k_c + \frac{1}{2} \epsilon^{ac} e_a e_c .
\label{U3Curvatures}
\end{eqnarray}

The NR action is
\begin{eqnarray}\label{U3Full-NRaction}
\nonumber  S_{NR}& =& \int  \tilde{\alpha}_1 \left[2k_a R^a(\omega^b)+ e_a R^a(e^b)-2m R(\tau )-2sR(k)-2t R(t)\right]\\
&&
 -\tilde{\alpha}_2\left[-2s R(\omega)+ \omega_a R^a(\omega^b)\right]  
+ 2\tilde{\alpha}_3\left[e^aR_a(\omega^b) -\tau R(s)-m R(\omega) \right].
\end{eqnarray}

Because the bilinear form does not become degenerate in the contraction process the CS formulation guarantees that the equation of motion from this NR action are the independent vanishing of all the curvatures (\ref{U3Curvatures}). Thus, the dynamics is not redundant.

\section{Discussion}  
\label{discussion}

We have examined a class of NR CS Maxwell gravities in 2+1 dimensions obtained as the NR particle limit of relativistic CS Maxwell gravity. At the Lagrangian level, the relativistic action written as a series expansion in terms of the dimensionless parameter $\xi$ in general contains infinities. The coefficient of the most divergent term is always NR invariant \cite{BGN} while the remaining terms, in particular the finite one, are in general not  invariant under the NR symmetry group. Following the idea of \cite{GO,BRZ} we consider the addition of an extra piece such that the total relativistic Lagrangian has a NR expansion in which the first term is finite. Inspired by \cite{Bergshoeff-Rosseel}, we consider the CS action associated with [Maxwell]$\ \oplus\ u(1)\oplus u(1)$ algebra in three-dimensional space-time. In this way, a finite NR CS Lagrangian is obtained, with an invariant bilinear form which is generically degenerate. Therefore the NR Maxwell gravity has field equations that do not determine all the dynamical fields.  We found two ways to cure this difficulty.

First, the field equations (\ref{degEOM}) themselves suggest various ways to cure the dynamical indeterminacy. One possibility that gives a Lagrangian invariant under NR transformations with a non-degenerate bilinear form corresponds to the vanishing of the central charge associated to the non commutativity of the Galilean boost, i. e. $[\tilde{G}_a,\tilde{G}_b]=0$. The Lagrangian depends on four parameters, $\tilde\alpha_1, \tilde\alpha_2, \tilde\alpha_0,\tilde\alpha_6$ and when the last three are put to zero, the theory becomes the NR limit of a CS relativistic system with [Maxwell]$\ \oplus\ u(1)$ gauge algebra.  This case is studied in subsection \ref{section-NR2}.

Second, starting with [Maxwell] $\ \oplus\ u(1)\oplus u(1)\oplus u(1)$ at the relativistic level it is shown in section \ref{section-NR3} that there is a generalization of the trasformation used in \cite{Bergshoeff-Rosseel} for this relativistic algebra that leads to a NR algebra admitting a non degenerate bilinear form. The NR CS action obtained through the contraction contains the Extended Bargmann Gravity and the Exotic Gravity as subcases.

One of the points that needs further study is to analyze in detail the action (\ref{U3Full-NRaction}) and its dynamical contents. In particular to construct the second order metric formulation, i. e., the NR theory obtained by substituting $\omega$ and $\omega^a$  into the NR action (\ref{U3Full-NRaction}). Note however that the advances in the dynamical analysis performed for the system of subsection \ref{section-NR2} apply for the system of section \ref{section-NR3} (vanishing of curvatures in (\ref{U3Curvatures})). In paticular the explicit solution for the connection fields $\omega$ and $\omega_a$ worked out in the appendix B are the same.

The torsion equations $R_a(e^b)=0$ can be algebraically solved for $\omega^a$ and it allows us to write the field equations as a second order system. However, if the torsion equation is not obtained as the variation of the action with respect to $\omega^a$, the second order system obtained by substituting $\omega^a$ in the action needs not to be equivalent to the first order one and in general possesses different degrees of freedom. In particular, this might be the case when the torsion equation is enforced by a Lagrange mutiplier, in which case the resulting second order system may have propagating degrees of freedom while the system with torsion not forced to vanish may have no propagating degrees of freedom at all \cite{DJT,Wise,BHT}. 

Another interesting problem is the construction of a supergravity theory associated to super Maxwell algebra \cite{Bonanos:2009wy} in 2+1 dimensions. 

Also in the context of supersymmetry, note that in this work we have reinforced the idea that adding bosonic fields at the relativistic level helps in having a well-defined NR limit. As shown for instance in \cite{Howe:1995zm,Bonanos:2009wy,Fuentealba:2017fck,Basu:2017aqn}, supersymmetry may require additional bosonic fields to be well-defined too. If the number and nature of the extra fields coincide in formulating the supersymmetric version for the theory and in having a NR limit it would suggest that there is a deeper connection. Then, this would be a criteria to select classical relativistic theories before addressing their quantization.

\section*{Acknowledgments}
We are grateful to Eric Bergshoeff, Niels Obers, and Jan Rosseel for insightful discussions and in particular, for pointing out to us the issue discussed in section 3.3. We are also grateful to Axel Kleinschmidt and Oscar Fuentealba for  enlightening discussions.  JG has been supported in part by FPA2013-46570-C2-1-P and Consolider CPAN and by the Spanish goverment (MINECO/FEDER) under project MDM-2014-0369 of ICCUB (Unidad de Excelencia Mar\'\i a de Maeztu). JG has also been visiting professor to the Universidad Austral de Chile under grant PAI801620047 from CONICYT. EF is partially funded by Fondecyt grant 11150467. DH and LA  are partially founded by Conicyt grants 21160649 and 21160827, respectively. JZ has been partially funded by Fondecyt grants numbers 1140155 and 1180368. The Centro de Estudios Cient\'ificos (CECs) is funded by the Chilean Government through the Centers of Excellence Base Financing Program of Conicyt. 

\appendix

\vspace{0.8cm}

{\Large {\bf Appendix}}

\section{Chern-Simons invariance}
Let us recall two important properties of CS actions \cite{HZ}. First, we show that the quasi-invariance under the symmetry group is guaranteed. The infinitesimal  variation of the Lagrangian under the gauge symmetry, $\delta_\lambda A=D\lambda=d\lambda+[A,\lambda]$,  is a total derivative
\be
\delta_\lambda\langle AdA+\frac{1}{3}A[A,A]\rangle = d \langle 2  F(A)\lambda - A \delta_\lambda  A \rangle,
\ee
where we used the Bianchi identity $DF=0$. Note that the CS Lagrangian requires an invariant bilinear form.

Second, we show that if the invariant bilinear form, $g_{AB}\equiv\langle X_A,X_B\rangle$, is non-degenerated then the CS curvature vanishes. And therefore, all curvature components which are themselves curvatures for each field, vanish too. The CS equation of motion is
\be
0=\langle F(A)\delta A\rangle=F^A\delta A^B g_{AB}.
\ee

Thus, because $\delta A^B$ is arbitrary if $g_{AB}$ is invertible we have $F^A=0$. Note however that the contrary is not neccesarily true:  The vanishing of the curvatures does not guarantee the existence of a non-degenerated invariant bilinear form.

Note that  to find an invariant bilinear form for a given algebra is a straightforward linear problem. The algebra is encoded in the structure constants $[X_A,X_B]=f_{AB}^{\ \ \  C}X_C$ and the invariance of a generic bilinear form $g_{AB}=\langle X_A,X_B\rangle$ is the statement $\langle X_A,X_B\rangle=\langle X_A+\delta X_A,X_B+\delta X_B\rangle$ or equivalently
\ba
\langle X_A,X_B \rangle-\langle X_A+[X_A,X_C], X_B+[X_B,X_C]\rangle &=&0 \\
f_{BC}^{\ \ \  D}g_{AD}+f_{AC}^{\ \ \  D}g_{DB}&=&0\label{ecua}
\ea

Then, given the structure constants the most general invariant bilinear form is determined. Weather that solution for $g_{AB}$ is degenerate or not is a different question. Note further that if $g_{AB}$ is non-degenerate the most general invariant bilinear allows us to have all the quadratic Casimir objects. In fact, the equation $[g^{AB}X_AX_B,X_C]=0$ is equivalent to (\ref{ecua}).

\section{Explicit $\omega$ and $\omega^a$ connections}

To compute the non-relativistic components of the connections, $\omega$ and $\omega_a$, we follow the standard strategy (see \cite{Andringa.PhD.Thesis}) but using our notation. \\
{\bf Computation of $\omega_\mu$:}\\
 Consider 
\begin{equation}
\frac{1}{2}R_{\mu \nu}^{a} (e^c) = \partial_{[\mu}e^{a}_{\nu ]}   +\epsilon^{ab} \tau_{[\mu} \omega_{b \nu ]}   + \epsilon^{ab}  \omega_{[\mu}  e_{b\nu]} = 0,
\end{equation}
then we multiply by $e_{a\rho}$, i.e., $\frac{1}{2}R_{\mu \nu}^{a}(e^c)e_{a\rho} =0$ and write its cyclic permutations
\begin{eqnarray}
\quad \partial_{[\mu}e^{a}_{\nu]}     e_{a\rho} + \epsilon^{ab} \tau_{[\mu} \omega_{b \nu]} e_{a\rho} + \omega_{\rho [\nu \mu]} & = & 0, \\
\quad \partial_{[\rho}e^{a}_{\mu]}     e_{a\nu} + \epsilon^{ab} \tau_{[\rho} \omega_{b \mu]} e_{a\nu} + \omega_{ \nu [\mu \rho]} & = & 0, \\
\quad \partial_{[\nu}e^{a}_{\rho]}     e_{a\mu} + \epsilon^{ab} \tau_{[\nu} \omega_{b \rho]} e_{a\mu} + \omega_{ \mu [\rho \nu]} & = & 0,
\end{eqnarray}
where $\omega_{\rho \nu \mu} \equiv \epsilon^{ab}\omega_{[\mu } e_{b |\nu]} e_{a \rho}$ is anti-symmetric in the first two and the last two indexes. We sum the first two equations and subtract the third one to get
\begin{equation}
\partial_{[\mu}e^{a}_{\nu]}     e_{a\rho} + \epsilon^{ab} \tau_{[\mu} \omega_{b \nu]} e_{a\rho}   + \partial_{[\rho}e^{a}_{\mu]}     e_{a\nu} + \epsilon^{ab} \tau_{[\rho} \omega_{b \mu]} e_{a\nu}   - \partial_{[\nu}e^{a}_{\rho]}     e_{a\mu} -\epsilon^{ab} \tau_{[\nu} \omega_{b \rho]} e_{a\mu} + \omega_{\rho \nu  \mu}=0.
\end{equation}

To invert the relation we have $\omega_{\rho \nu \mu}e^{b \nu} e^{a\rho} = \epsilon^{ab}\omega_\mu $, then
\begin{multline}
-\epsilon^{ab}\omega_\mu = \frac{1}{2} (  \partial_\mu e^{a}_{\nu} e^{b\nu} - \partial_\nu e^{a}_{\mu} e^{b\nu}    )+ \frac{1}{2} \epsilon^{ab} \tau_\mu \omega_{b\nu} e^{b\nu} \\+ \frac{1}{2}  ( \partial_\rho e^{b}_{\mu} e^{a\rho} - \partial_\mu e^{b}_{\rho} e^{a\rho}  ) - \frac{1}{2} \epsilon^{bc}\tau_\mu \omega_{c\rho} e^{a\rho} - \partial_{[\nu}e_{\rho]}^{c}e_{c\mu} e^{b\nu} e^{a\rho},
\end{multline}
we used $\tau^\mu e_{\mu a} =0 $.  Now, multiplying by $-\epsilon_{ab}/2$ (remember $\epsilon_{ab}\epsilon^{ab}=-2$), it yields
\begin{equation}\label{omega}
\omega_\mu  = \epsilon_{ab} \partial_{[\nu} e^{a}_{\mu]} e^{b\nu} + \frac{1}{2} \epsilon_{ab}\partial_{[\nu} e^{c}_{\rho]} e_{c\mu} e^{a\nu} e^{b\rho}- \frac{1}{2} \tau_{\mu} \omega_{b\nu} e^{b\nu} .
\end{equation}
{\bf Computation of $\omega_{\mu}^{a}$:}\\ 

First, we replace the expression \eqref{omega} into the curvature $R_{\mu \nu}^{a}(e^b)$ and compute the expression
\be
e^{\mu c}\tau^{\nu} R_{\mu \nu}^{a}(e^{b}) + e^{\mu a}\tau^{\nu} R_{\mu \nu}^{c}(e^{b}) =0,
\ee
this gives the symmetric part
\begin{equation}\label{constrain-1}
\epsilon^{(a|b}e^{\mu |c)} \omega_{b\mu} = 2 e^{\mu( a} \tau^{\nu} \partial_{[\mu} e_{\nu]}^{c)}.
\end{equation}

Now we use the $R (m)=0$ equation.  By computing $2R_{\mu \nu} (m) e^{a\mu}e^{b\nu} = 0$ we obtain the antisymmetric part
\begin{equation}\label{constraint-2}
\epsilon^{[a|b} e^{\mu |c]} \omega_{b\mu} = - e^{[a\mu} e^{c]\nu} \partial_{[ \mu} m_{\nu]},
\end{equation}

From the curvature equation we also need the contraction $2R_{\mu \nu} (m)e^{a\mu} \tau^{\nu} = 0$, it yields
\begin{equation}\label{constraint-3}
\epsilon^{ab} \tau^{\mu} \omega_{b\mu} =-2 e^{a\mu} \tau^{\nu} \partial_{[\mu} m_{\nu]}.
\end{equation}

Summing \eqref{constrain-1} and \eqref{constraint-2}, and then multiplying by $e_{\sigma c}$ and $-\frac{\epsilon_{ab}}{2}$,  it yields
\begin{equation}
\omega_{\sigma a} = \frac{\epsilon_{ab}}{2} \left( \tau^{\mu} \partial_{[\sigma} e^{b}_{\mu]} +e^{b\mu} e_{c\sigma} \tau^\nu \partial_{ [\mu} e_{\nu ]}^{c} +e^{b\mu} \partial_{[\sigma} m_{\mu]} +\tau_{\sigma} e^{b\nu} \tau^{\mu} \partial_{[\mu} m_{\nu]}\right), 
\end{equation}
we used the relation $	e^{\mu c} e_{\sigma c} = \delta^{\mu}_{\sigma}-\tau^{\mu}\tau_{\sigma}$ and the \eqref{constraint-3}.

	


\end{document}